\documentclass[12pt]{article}
\usepackage{amsmath}
\usepackage{amssymb}
\begin{document}

\thispagestyle{plain} \setcounter{page}{1}

\title{Higher order first integrals of motion in a gauge covariant 
Hamiltonian framework}

\author{Mihai Visinescu \thanks{E-mail:~~~mvisin@theory.nipne.ro}\\
{\small \it Department of Theoretical Physics,}\\
{\small \it National Institute for Physics and Nuclear Engineering,}\\
{\small \it P.O.Box M.G.-6, Magurele, Bucharest, Romania}}
\date{}

\maketitle

\begin{abstract}

The higher order symmetries are investigated in a covariant Hamiltonian 
formulation. The covariant phase-space approach is extended to include the 
presence of external gauge fields and scalar potentials. The special 
role of the Killing-Yano tensors is pointed out. Some non-trivial 
examples involving Runge-Lenz type conserved quantities are explicitly 
worked out.

~

Pacs 11.30.-j, 11.30.Ly, 11.30.Na

~

\end{abstract}

\section{Introduction}

In a recent paper van Holten \cite{JWH} proposed a technique for 
deriving conserved quantities in a covariant formalism of the dynamics 
of particles in external gauge fields. Using a completely covariant 
phase-space formulation, he studied a set of generalized Killing 
equations in order to produce constants of motion in a covariant 
way. This procedure applies to conserved quantities which are higher 
order polynomials in the momenta, as well as the spinning particle 
models in curved space-time, involving Grassmann variables to take into 
account fermionic degrees of freedom \cite{GRH}. Van Holten's algorithm 
was used successively to construct conserved quantities in a 
non-Abelian monopole field \cite{HN} and on generalized Euclidean 
Taub-NUT space \cite{Ng}.

The aim of this paper is to extend the technique from \cite{JWH} for 
the dynamics of particles in external gauge fields and scalar 
potentials. The inclusion of scalar potentials permits us to extend the 
applicability of the covariant approach to more complex cases. For 
example, the motion in a Kepler-Coulomb (KC) potential with Runge-Lenz 
(RL) type conserved quantities is affected in a non-trivial way by the 
external gauge fields. Moreover, some generalizations of the KC systems 
have interesting integrability properties connected with the existence 
of additional hidden integrals of motion which are polynomial functions 
in the momenta \cite{JTH,VE,BEHR,BH}. It is interesting to investigate 
the superintegrability of the generalized KC systems on $N-$dimensional 
curved spaces in conjunction with external gauge fields.

In the absence of gauge fields, the system of generalized Killing 
equations separates into two groups: one group involves the terms of 
the integral of motion of odd degree in the momenta and the other 
involves only the terms of even degree in the momenta \cite{JTH}. In 
the presence of gauge fields, such a separation is not possible and the 
analysis of the system of differential equations is more intricate. A 
few examples will explicitly illustrate the complexity of the systems 
of conditions for the integrals of motion.

In general, the explicit and hidden symmetries of a space-time are 
encoded in the multitude of Killing vectors and higher order St\" 
ackel-Killing (SK) tensors respectively. Another natural generalization 
of the Killing vectors is represented by the Killing-Yano (KY) tensors 
\cite{KY}. A KY tensor generates additional supercharges in the dynamics 
of pseudo-classical spinning particles, realizing a natural connection 
with supersymmetries \cite{GRH}. Passing to quantum Dirac equation it 
was discovered \cite{CML} that KY tensors generate non-standard Dirac 
operators which commute with the standard one. These two 
generalizations of the Killing vectors could be related. In some cases 
a SK tensor is associated with a KY tensor, namely a rank $2$ SK
tensor could be written as a product of KY tensors.

The role of KY tensors with the framework extended to include 
electromagnetic interactions was pointed out by Tanimoto \cite{MT}. It 
was found the necessary condition of the electromagnetic field 
$F_{\mu\nu}$ to maintain the supersymmetry generated by a KY tensor. In 
this paper we retrace the argument in \cite{MT} and see the role of KY 
tensors in connection with the covariant Hamiltonian framework.

The plan of the paper is as follows. In Section 2 we establish the 
generalized Killing equations in a covariant framework including 
external gauge fields and scalar potentials. In Section 3 we discuss 
the role of KY tensors using the condition on the electromagnetic 
tensor $F_{\mu\nu}$ from \cite{MT}. In the next Section we produce some 
non-trivial examples for a KC system in the presence of external 
fields: constant electric field, spherically symmetric magnetic fields 
and a magnetic field along a fixed direction.
Finally, Section 5 is devoted to conclusions.

\section{Conditions for a conserved quantity}

Let $(\mathcal{M}, \mathbf{g})$ denote a $N$-dimensional 
manifold with the metric tensor $\mathbf{g}$. The manifold 
$\mathcal{M}$ is usually the space-time manifold or its Euclidean 
version, i.e. a Riemannian manifold of dimension $4$. However in many 
modern physical applications, such as superspace, Kaluza-Klein models 
and string theories, the manifold $\mathcal{M}$ can be an arbitrary 
manifold.

The classical dynamics of 
a point charge $q$ of mass $M$ in the external Abelian gauge field $A_i$ 
and a scalar potential $V(x^i)$ with respect to a system of local position 
coordinates $x^i$ is described by the Hamiltonian:
\begin{equation}\label{H}
H = \frac{1}{2M} g^{ij} (p_i - q A_i)  (p_j - q A_j) + V\,.
\end{equation}
Here $g^{ij}$ are the components of the contravariant metric on 
$\mathcal{M}$ and $p_i$ are the canonical momenta conjugate to the 
coordinates $x^i$. We also adopt the notation $~_{,i}$ for partial 
differentiation with respect to $q^i$ and $~_{;i}$ for the covariant 
derivative constructed with the Levi-Civita connection on $\mathcal{M}$ 
with respect to the metric $\mathbf{g}$. We shall also employ the usual 
summation convention over the repeated upper and lower indices from 
$1$ to $N$.

The equations of motion can be written with the use of Poisson bracket
\begin{equation}\label{PB}
\{f,g\} = \frac{\partial f}{\partial x^i} \frac{\partial g}{\partial p_i} 
-\frac{\partial f}{\partial p_i} \frac{\partial g}{\partial x^i} \,.
\end{equation}

The disadvantage of this approach is that the canonical momenta $p_i$ 
and implicitly the Hamilton equations of motion are not manifestly 
gauge covariant. Using van Holten's receipt \cite{JWH}, this drawback 
can be removed introducing the gauge invariant momenta
\begin{equation}\label{Pi}
\mathbf{\Pi} = \mathbf{p} - q\mathbf{A} = M \dot{\mathbf{x}} \,.
\end{equation}

The Hamiltonian becomes
\begin{equation}\label{HPi}
H = \frac{1}{2M} g^{ij} \Pi_i  \Pi_j + V\,,
\end{equation}
and the equations of motion are derived using the covariant Poisson 
brackets \cite{GPS}
\begin{equation}\label{covPB}
\{f,g\} = \frac{\partial f}{\partial x^i} \frac{\partial g}{\partial 
\Pi_i} -\frac{\partial f}{\partial \Pi_i} \frac{\partial g}{\partial x^i} 
+ q F_{ij}\frac{\partial f}{\partial \Pi_i} 
\frac{\partial g}{\partial \Pi_j} \,.
\end{equation}
where $F_{ij} = A_{j;i} - A_{j;i}$ is the field strength.

The fundamental Poisson brackets are
\begin{equation}
\{ x^i , x^j \} = 0\,, \; \{ x^i , \Pi_j \} = \delta^i_j \,,\;
\{ \Pi_i , \Pi_j \} = q F_{ij}\,,
\end{equation}
showing that the momenta $\mathbf{\Pi}$ are not canonical. A direct 
computation of the Hamilton's equations gives:
\begin{subequations}
\begin{eqnarray}
\dot{x^i} = \{ x^i, H\} = \frac{1}{M} g^{ij} \Pi_j\,, \\
\dot{\Pi_i} = \{ \Pi_i, H\} = q F_{ij} \dot{x^j} 
- V_{,i}\,. 
\end{eqnarray}
\end{subequations}

In terms of phase-space variables $(x^i , \Pi_i)$ the conserved 
quantities of motion read:
\begin{equation}
K = K_0  + \sum^{p}_{n=1}\frac{1}{n!} K_n^{i_1 \cdots i_n} (x) 
\cdots \Pi_{i_1} \Pi_{i_n}\,,
\end{equation}
where $K_n^{i_1 \cdots i_n}\,,\; n=1,\cdots p$ are contravariant 
tensors on $\mathcal{M}$ taken to be completely symmetric.
Its bracket with the Hamiltonian vanishes $\{K, H\} = 0$  and this 
yields the series of constraints:
\begin{subequations}\label{constr}
\begin{eqnarray}
&& K_1^i V_{,i} = 0\,, \label{1}\\
&& K_{0,i} +  q F_{ji} K_1^j = M K_{2 i}^j V_{,j}\,. \label{0}\\
&& K_n^{(i_1 \cdots i_n;i_{n+1})} + q F_j^{~(i_{n+1}} 
K_{n+1}^{i_1 \cdots i_n) j}
= \frac{M}{(n+1)}  K_{n+2}^{i_1 \cdots i_{n+1}j} V_{,j} \label{n}\,,
\nonumber\\ 
&& ~~~~~~~~~~~~~~~~~~~~~~~~~~~~~~~~~~~~~~~~~~~~ 
 for ~~ n= 1\,,\cdots (p-2)\,,\\
&& K_{p-1}^{(i_1 \cdots i_{p-1};i_p)} + q F_j^{~(i_p} 
K_p^{i_1 \cdots i_{p-1}) j} =0 \,,\label{p-1}\\
&&K_p^{(i_1 \cdots i_p;i_{p+1})} =0\,. \label{p}
\end{eqnarray}
\end{subequations}
Here the parentheses denote full symmetrization over the indices 
enclosed.

Examining the above hierarchy of constraints \eqref{constr} we remark 
that in the absence of the gauge field strength $F_{ij}$, equations 
\eqref{constr} separate into two groups. One group involves the terms 
of $K$ of odd degree in the momenta and the other involves only the 
terms of $K$ of even degree in the momenta \cite{JTH}. Here, the 
presence of the gauge field strength $F_{ij}$ mixes up the terms of $K$ 
of even and odd degrees in the momenta and consequently the system of 
coupled equations \eqref{constr} is more intricate. Moreover, only the 
last equation  \eqref{p} for the leading order term 
$K_p^{i_1 \cdots i_p}$ defines the component of a SK tensor of rank 
$p$. Note that again, in the absence of the field strength $F_{ij}$,
equation \eqref{p-1} also defines a SK tensor 
$K_{p-1}^{i_1 \cdots i_{p-1}}$, but here this is not the case.

\section{Role of KY tensors}

The next most simple objects that can be studied in connection with the 
symmetries of a manifold $(\mathcal{M}, \mathbf{g})$ after SK tensors 
are KY tensors \cite{KY}. Their physical utility remained unclear until 
Floyd \cite{RF} and Penrose \cite{RP} showed that a SK tensor of rank 
$2$ of the $4$-dimensional Kerr-Newman space-time admits a certain 
square root which defines a KY tensor. On the other hand it was 
realized \cite{GRH} that a KY tensor generates additional supercharges 
in the dynamics of pseudo-classical spinning particles. In the quantum 
framework it was shown \cite{CML} that KY tensors produce 
conserved non-standard Dirac type operators which commute with the 
standard one.

A differential $p$-form $f$ is called a KY tensor if its covariant 
derivative $f_{\mu_1 \cdots \mu_p ; \lambda}$ is totally antisymmetric. 
Equivalently, a tensor is called a KY tensor of rank $p$ if it is 
totally antisymmetric and satisfies the equation
\begin{equation}\label{KY}
f_{\mu_1 \cdots (\mu_p ; \lambda)} = 0 \,.
\end{equation}

These two generalizations SK \eqref{p} and KY \eqref{KY} of the Killing 
vectors could be related. Let $f_{\mu_1 \cdots \mu_p}$ be a KY tensor, 
then the tensor field
\begin{equation}\label{K2ff}
K_{2\mu\nu} = f_{\mu \mu_2 \cdots \mu_p}
f^{\mu_2 \cdots \mu_p}_{~~~~~~~\nu} \,,
\end{equation}
is a SK tensor and one sometimes refers to it as the associated tensor 
with $f$.

A typical example is represented by the Euclidean Taub-NUT space which 
admits a RL vector whose components are SK tensors. These SK 
tensors can be written as symmetrized products of KY tensors \cite{VV}. 
An important physical consequence of this possibility to decompose a 
SK tensor in terms of antisymmetric KY tensors is the absence of 
gravitational anomalies \cite{BC,CMV}.

The role of KY tensors in the motion of pseudo-classical spinning point 
particles was extended by Tanimoto \cite{MT} to include electromagnetic 
interactions. He obtained the condition of the electromagnetic field 
$F_{\mu\nu}$ to maintain the non generic supersymmetry associated with 
a KY tensor $f$ of rank $p$. This condition can be expressed as
\begin{equation}\label{condp}
F_{\nu[\mu_p}f_{\mu_1 \cdots \mu_{p-1}]}^{~~~~~~~~~\nu} = 0\,,
\end{equation}
where the indices in square brackets are to be antisymmetrized.
In particular, for a Killing vector $K_1^\nu$ this conditions is
\begin{equation}\label{cond1}
F_{\mu\nu} K_1^\nu =0 \,.
\end{equation}

In what follows we shall investigate the consequences of this 
condition for the series of constraints \eqref{constr}. To be more 
definite, we shall limit ourselves to the first three constraints 
\eqref{1} - \eqref{n} for $n=1$, assuming that the SK tensor 
$K_{2\mu\nu}$ is associated with a KY tensor $f_{\mu\nu}$
\begin{equation}\label{Kff}
K_{2\mu\nu} = f_{\mu\lambda}f^{\lambda}_{~\nu}\,.
\end{equation}
In this case, condition \eqref{condp} for the electromagnetic field 
$F_{\mu\nu}$ reads
\begin{equation}\label{cond2}
F_{\lambda[\mu}f_{\nu]}^{~\lambda } = 0\,.
\end{equation}

Using the antisymmetric properties of the KY tensors and 
electromagnetic field $F_{\mu\nu}$, for a  SK tensor of the form 
\eqref{Kff} in conjunction with condition \eqref{cond2} we get that 
in the l. h. s. of equation \eqref{n} the 
term $q F_j^{~~i_2}K_2^{i_1 j}$ vanishes. Consequently we have a 
relation between the odd terms $K_1, K_3$ as in the absence of the 
electromagnetic field. The same argument applies to equation \eqref{0} 
where condition \eqref{cond1} implies a relation only between even 
terms $K_0, K_2$.

In conclusion, condition \eqref{condp} which plays an important role in 
the construction of superinvariants for the motion of pseudo-classical 
spinning char\-ged point particles proves to produce significant
simplifications in the series of constraints \eqref{constr} for the 
higher order integrals of motion.

\section{Explicit examples}

Let us illustrate these general considerations by some non trivial
examples. In what follows we consider $\mathcal{M}$ to be a 
$3$-dimensional Euclidean space $\mathbb{E}^3$ and in these 
circumstances it is more convenient to get rid of a difference between 
covariant and contravariant indices. 

We are looking for  constants of motion of the form
\begin{equation}\label{RLE}
K = \frac{1}{2} K_{2ij} \Pi_i \Pi_j + K_{1i} \Pi_i + K_0 \,.
\end{equation}

Recently the three-dimensional integrable systems were investigated 
assuming that there are additional integrals with quadratic dependence 
of momenta \cite{GNP1,GNP2}. In some cases the potentials are 
compatible with complete integrability and then one gets separation of 
variables. In the Hamiltonian-like constants the role 
of the metric is played by the SK tensor $K_{2ij}$ and the 
corresponding potential $K_{0}$ is related to the original one $V$ as 
in \eqref{0}. The reciprocal relation between the standard metric $g_{ij}$ 
and SK tensor $K_{2ij}$ has a geometrical interpretation \cite{RH}: it 
implies that if $K_{2}^{ij}$ are the contravariant components of a SK 
tensor with respect to the inverse metric $g^{ij}$, then $g_{ij}$ must 
represent a SK tensor with respect to the inverse metric defined by  
$K_{2}^{ij}$.

In what follows we investigate the constant of motion in a KC potential 
adding different types of electric and magnetic fields. To put in a concrete 
form, we consider the motion of a point charge $q$ of mass $M$ in 
the Coulomb potential $Q/r$ produce by a charge $Q$ when some external 
electric or magnetic fields are also present.

\subsection{Constant electric field}

In a first example we consider the electric charge $q$ moving in the 
Coulomb potential with a constant electric field $\mathbf{E}$ present. 
Therefore in the potential $V(x^i)$ we include the Coulomb potential 
and $\mathbf{E}\cdot\mathbf{r}$ for the external electric field. The 
corresponding Hamiltonian is:
\begin{equation}\label{Hel}
H = \frac{1}{2 M} \mathbf{\Pi}^2 + q\frac{Q}{r} - 
q \mathbf{E}\cdot\mathbf{r}\,,
\end{equation}
with $\mathbf{\Pi} = M \dot{\mathbf{r}}$ in spherical coordinates of 
$\mathbb{E}^3$.

As it is known the non relativistic KC problem admits two vector 
constants of motion, namely the angular momentum
\begin{equation}\label{am}
\mathbf{L} =\mathbf{r} \times \mathbf{\Pi}\,,
\end{equation}
and the RL vector 
\begin{equation}\label{RL}
\mathbf{K} = \mathbf{\Pi} \times \mathbf{L} + 
MqQ\frac{\mathbf{r}}{r} \,.
\end{equation}

The components $K_{2ij}$ of the constant of motion \eqref{RLE} are SK tensors, 
satisfying equation \eqref{p} for $p = 2$. For the KC problem it proves
adequate to choose for the SK tensor of rank $2$ the simple form 
\cite{MC}
\begin{equation}\label{K2}
K_{2ij} = 2 \delta_{ij} \mathbf{n}\cdot \mathbf{r} - (n_i r_j + n_j n_i)\,,
\end{equation}
written in spherical coordinates with $\mathbf{n}$ an arbitrary 
constant vector.

In the presence of a constant electric field $\mathbf{E}$ it proves 
convenient to choose $\mathbf{n}$ along $\mathbf{E}$ and we start 
to solve the hierarchy of constraint \eqref{constr} with a solution 
of equation \eqref{p} of the form \eqref{K2} with 
$\mathbf{n}=\mathbf{E}$. Using this form for $K_{2ij}$ and the 
derivative of the potential $V$ corresponding to the Hamiltonian
\eqref{Hel}
\begin{equation}\label{Vi}
V_{,i} = - \frac{qQ}{r^3} r_i - q E_i\,,
\end{equation}
we get from \eqref{0} after a straightforward calculation 
\begin{equation}\label{K0}
K_0 = \frac{MqQ}{r} \mathbf{E}\cdot \mathbf{r} - 
\frac{Mq}{2} \mathbf{E}\cdot [\mathbf{r} \times 
(\mathbf{r} \times \mathbf{E})]\,.
\end{equation} 

Concerning equation \eqref{1} with the derivative of the potential 
\eqref{Vi}, it is automatically satisfied with a vector $\mathbf{K}_1$ 
of the form
\begin{equation}\label{K1}
\mathbf{K}_1 = \mathbf{r} \times \mathbf{E}\,,
\end{equation}
modulo an arbitrary constant factor. This vector 
$\mathbf{K}_1$
contribute to a conserved quantity with a term proportional to the 
angular momentum $\mathbf{L}$ along the direction of the electric field 
$\mathbf{E}$.

In conclusion, when a uniform constant electric field is present, the 
KC system admits two constants of motion $\mathbf{L}\cdot\mathbf{E}$ 
and $\mathbf{C}\cdot\mathbf{E}$ where $\mathbf{C}$ is a generalization 
of the RL vector \eqref{RL} (see also \cite{Red}):
\begin{equation}\label{RLel}
\mathbf{C} = \mathbf{K} - \frac{Mq}{2}
\mathbf{r} \times ( \mathbf{r} \times \mathbf{E})\,.
\end{equation}

For this system with two additional constants of motion, the separation 
of variables is possible as it was demonstrated in \cite{GNP1,GNP2}. 
Here we confine ourselves to mention that the separation of variables 
for this definite problem can be found in many textbooks. For example, 
L. D. Landau and E. M. Lifshitz \cite{LL} have given an expression in 
{\it parabolic coordinates} for the constant of motions for the KC 
problem plus a constant electric field.

\subsection{Spherically symmetric magnetic field}

In what follows we consider an external spherically symmetric magnetic 
field
\begin{equation}\label{Bs}
\mathbf{B} = f(r) \mathbf{r}\,,
\end{equation}
and the Coulomb potential acting on a electric charge $q$.

Spherically symmetric magnetic fields appear in many interesting 
physical problems, the most notable configuration being the Dirac 
charge-monopole system. Many different formulations of the 
charge-monopole system have been well discussed in the literature, 
see e.g. \cite{MP}.

Here we are interested in higher order constants of motion for 
spherically symmetric magnetic configurations involving a SK tensor of 
rank $2$. Again we truncate the system \eqref{constr} taking 
$K_p^{i_1 \cdots i_p} = 0$ for all $p \geq 3$. For the beginning the 
scalar function $f(r)$ is not fixed, its form will be determined from 
the hierarchy of constraints \eqref{constr}. For $K_{2ij}$ we use the 
form \eqref{K2} typical for the SC system with spherical symmetry. 
Equation \eqref{n} for $n=1$ with
\begin{equation}
F_{ij} = \epsilon_{ijk} B_k = \epsilon_{ijk} r_k f(r) \,,
\end{equation}
is
\begin{equation}
K_{1(i,j)} = - q f(r) [ (\mathbf{n} \times \mathbf{r})_{(i} r_{j)}]\,,
\end{equation}
with $\mathbf{n}$ an arbitrary unit constant vector. It is easy to get 
for the vector $\mathbf{K}_1$ the solution
\begin{equation}
K_{1i} = q \left [\int r f(r) dr\right ] (\mathbf{n} \times 
\mathbf{r})_i \,,
\end{equation}
and equation \eqref{1} is obviously satisfied.

For $K_0$, equation \eqref{0} can be solely solved making choice of 
a definite form for the function $f(r)$
\begin{equation}
f(r) = \frac {g}{r^{5/2}}\,,
\end{equation}
with
$g$ a constant connected with the strength of the magnetic field.
For this function $f(r)$ the energy of the magnetic field diverges at
$r = 0$ and $r \to \infty$. Of course such a special magnetic field 
\eqref{Bs} could be prepared only in a finite region of space and all 
present considerations are limited to this space domain.

With this special form of the function $f(r)$ we get
\begin{equation}
K_0 = \left [ \frac{MqQ}{r} - \frac{2g^2 q^2}{r}\right ](\mathbf{n}\cdot 
\mathbf{r})\,,
\end{equation}
and 
\begin{equation}
K_{1i} = - \frac{2gq}{r^{1/2}}(\mathbf{r}\times \mathbf{n})_i\,.
\end{equation}

Collecting the terms $K_0, K_{1i}, K_{2ij}$ the constant of motion 
\eqref{RLE} becomes
\begin{equation}
K = \mathbf{n}\cdot \left (\mathbf{K} + \frac{2gq}{r^{1/2}}\mathbf{L}
 - 2 g^2 q^2 \frac{\mathbf{r}}{r}\right )\,,
\end{equation}
with $\mathbf{n}$ an arbitrary constant unit vector and $\mathbf{K}, 
\mathbf{L}$ given by \eqref{RL}, \eqref{am} respectively. In contrast 
with the pure Coulomb potential, the presence of a spherically magnetic 
field prevents the separate conservation of the angular momentum 
$\mathbf{L}$ \cite{JWH}.

\subsection{Magnetic field  along a fixed direction}

Let us consider a magnetic field directed along a fixed unit vector 
$\mathbf{n}$
\begin{equation}\label{Bm}
\mathbf{B} = B(\mathbf{r} \cdot \mathbf{n}) \mathbf{n}\,,
\end{equation}
where, for the beginning, $B(\mathbf{r} \cdot \mathbf{n})$ is an 
arbitrary function.

As in the previous example we truncate the hierarchy of constraints 
\eqref{constr} for $p\geq 3$. Using \eqref{Bm} to evaluate the 
electromagnetic field strength $F_{ij}$ and choosing for $K_{2ij}$ the 
form \eqref{K2} with the arbitrary vector $\mathbf{n}$ along the 
magnetic field, equation \eqref{p} for $p = 1$ reads
\begin{equation}
K_{1(i,j)} = - q B (\mathbf{r} \times \mathbf{n})_{(i} n_{j)}\,.
\end{equation}
$K_{1i}$, solution of this equation, must satisfy also equation 
\eqref{1} and after some straightforward calculations we get
\begin{equation}
K_{1i} = q \left [\int r B(\mathbf{r}\cdot \mathbf{n}) 
d(\mathbf{r}\cdot \mathbf{n}) \right ] (\mathbf{r} \times 
\mathbf{n})_i \,.
\end{equation}

Equation \eqref{0} for $K_0$ proves to be solvable for a 
particular form of the magnetic field
\begin{equation}
\mathbf{B} = \frac{\alpha}{\sqrt{\alpha \mathbf{r} \cdot \mathbf{n} + \beta 
}}\; \mathbf{n}\,,
\end{equation}
with $\alpha$ and $\beta$ two arbitrary constants.
 
Finally  we get for $K_0$ and $K_{1i}$
\begin{eqnarray}
K_0 &=& \frac{MqQ}{r} (\mathbf{r}\cdot \mathbf{n}) +
\alpha q^2 (\mathbf{r} \times \mathbf{n})^2 \,,\\
K_{1i} &=& - 2 q \sqrt{\alpha \mathbf{r} \cdot \mathbf{n} + \beta }\;
(\mathbf{r} \times \mathbf{n})_i\,.
\end{eqnarray}

With these solutions, the constant of motion \eqref{RLE} for this 
configuration of the magnetic field superposed on the Coulomb potential 
becomes:
\begin{equation}\label{K}
K= \mathbf{n} \cdot \left [ \mathbf{K} + 2 q\sqrt{\alpha \mathbf{r} \cdot 
\mathbf{n} + \beta }\;\;\mathbf{L} \right ] + 
\alpha q^2 (\mathbf{r} \times \mathbf{n})^2 \,.
\end{equation}

As in the previous example the angular momentum $\mathbf{L}$ is no 
longer conserved, forming part of the constant of motion $K$ \eqref{K}.

\section{Concluding remarks}

In this paper we presented the hierarchy of constraints for integrals 
of motion in a gauge covariant Hamiltonian framework in the presence of 
Abelian gauge fields and scalar potentials. The formalism is valuable 
for the construction of the higher order constants of motions and 
deserves further studies.

An obvious extension is represented by the non-Abelian dynamics using 
the appropriate Poisson brackets \cite{JWH, HN}. On the other hand it 
was observed that the RL-type vector plays a role in the non-linear 
supersymmetry of various systems: fermion-monopole system 
\cite{MP1,MP2}, generalization of the Landau problem for 
non-relativistic electron coupled to electric and magnetic fields that 
produce a $1D$ crystal \cite{MP3}, reflectionless Poschl-Teller system 
in the context of the AdS/CFT holography and Aharonov-Bohm effect 
\cite{MP4}.

In order to exemplify the general considerations we worked out some 
simple examples in an Euclidean $3$-dimensional space and restricted to 
SK tensors of rank $2$. In a forthcoming paper \cite{MVprep} we shall 
illustrate the covariant Hamiltonian dynamics with more elaborate 
examples working on a $N$-dimensional curved space and involving higher 
ranks of SK tensors \cite{BEHR,BH}.

We concentrated on classical analysis and in principle 
there are no obstacles in passing to quantum mechanics. The results and 
their derivations are valid quantum mechanically, provided care is 
taken with the order of operators.

\subsection*{Acknowledgments}

The author would like to thank
M. S. Plyushchay for interest and interesting remarks concerning 
non-linear supersymmetries.
This work is supported by the CNCSIS Program IDEI - 571/2009, Romania.

\end{document}